\documentclass[a4paper]{article}

\usepackage[english]{babel}
\usepackage[utf8x]{inputenc}
\usepackage[T1]{fontenc}

\usepackage[a4paper,top=3cm,bottom=2cm,left=3cm,right=3cm,marginparwidth=1.75cm]{geometry}

\usepackage{amsmath}
\usepackage{graphicx}
\usepackage[colorinlistoftodos]{todonotes}
\usepackage[colorlinks=true, allcolors=blue]{hyperref}
\DeclareMathAlphabet{\pazocal}{OMS}{zplm}{m}{n}

\title{Rank and Rate: \\ Multi-task Learning for Recommender Systems\footnote{The paper was accepted to the 12th ACM Conference on Recommender Systems (RecSys '18)}}
\author{
Guy Hadash$^{*1}$,
Oren Sar Shalom$^{*2}$, 
Rita Osadchy$^{*3}$ \\
$^1$IBM Research, Haifa, Israel \\
$^2$Intuit, Hod HaSharon, Israel \\
$^3$University of Haifa, Haifa, Israel \\
\tt guyh@il.ibm.com, 
\tt oren.sarshalom@gmail.com,
\tt rita@cs.haifa.ac.il\\
}

\begin{document}
\maketitle

\begin{abstract}
The two main tasks in the Recommender Systems domain are the ranking and rating prediction tasks. The rating prediction task aims at predicting to what extent a user would like any given item, which would enable to recommend the items with the highest predicted scores. The ranking task on the other hand directly aims at recommending the most valuable items for the user. 
Several previous approaches proposed learning user and item representations to optimize both tasks simultaneously in a multi-task framework.
In this work we propose a novel multi-task framework that exploits the fact that a user does a two-phase decision process - first decides to interact with an item (ranking task) and only afterward to rate it (rating prediction task).

We evaluated our framework on two benchmark datasets, on two different configurations and showed its superiority over state-of-the-art methods.

\end{abstract}

\maketitle

\section{Introduction}
With the flourishing amount of products and information available on the web, recommender systems have become a prominent and useful tool. The benefits of personalized recommendations are unquestionable, hence the growing attention in this research topic. There are two main approaches for generating recommendations using Collaborative Filtering (\texttt{CF}) methods: rating prediction task and ranking task. At first, the dominant approach was the former. This task, encouraged by the Netflix Prize ~\cite{bennett2007netflix}, aims at predicting the rating a user would assign to given items. Given the predicted ratings that a user would assign to all items in the catalog, the recommended list can be composed of the items with the highest expected ratings. However, this approach introduces two major disadvantages: 
\begin{enumerate}
\item It does not prioritize the head of the recommended list \cite{cremonesi2010performance}. Since the final objective is to suggest interesting and valuable items for users, a small prediction error in the low rated items is not equivalent to a small prediction error in the  high rated items.
\item This task can be done only on rated items. As such it requires logged data of explicit ratings, which is much sparser than implicit feedback. Additionally, it suffers from selection bias, as user's decision not to interact with certain items is ignored \cite{marlin2009collaborative}.
\end{enumerate}
Therefore, in recent years more focus is given to the ranking task (also known as the selection task or the top-k recommendation problem), which is considered to better reflect user's needs ~\cite{gunawardana2009survey}. The goal of the ranking task is to compute rank scores, which may not predict the assigned rating, but rather they are directly used to generate the list of recommended items. 

Although the ranking task could be considered to be more important than the rating task, each of them encodes different bits of information and they may complete each other. Therefore, excelling in the secondary task may improve users and items representations and to lead to better performance on the primary task. This intuition paved the way for several frameworks that perform multi-task learning \cite{shi2013unifying,li2017one,li2016exploiting,li2016unified}.

In this paper, we refine the aforementioned intuition and argue that these two problems are not completely separate, but are part of a single process. A user first decides to interact with an item and subsequently decides on the explicit rating. We therefore design a multi-task learning approach that captures this two-phase decision process, and present its improved performance.

The rest of the paper is organized as follows. We start by giving the background (Section~\ref{sec:background}) needed for problem formulation. We then review the related works in Section \ref{sec:related}. We present our novel approach in Section~\ref{sec:ours}. Section~\ref{sec:results} includes the evaluation of the proposed method and the empirical comparison with the related work. Finally, Section~\ref{sec:conclusions} concludes the paper.

\section{Background} \label{sec:background}
\emph{Collaborative Filtering} (\texttt{CF}) methods receive as input a usage matrix $D$ consisting of $N$ users and $M$ items. Matrix $D$ may contain explicit ratings $r_{ui}$ that accompany the historical interactions.

\emph{Latent factor models} for \texttt{CF} represent each user $u$ and item $i$ by some low dimensional vectors $p_u$ and $q_i$ correspondingly. The predicted score of a given user-item pair $(u,i)$ is determined by: $s_{ui}=p_u^T\cdot q_i$. This score can be used for either ranking or rating prediction. Ranking algorithms usually learn user bias terms $b_u$, which are added to the predicted scores. Rating prediction algorithms may add to the predicted score the user and item biases, alongside with the mean rating.

The various algorithms differ in their objective function and means to generate the internal representation for users and items.
Such models for the rating prediction task are \texttt{SVD}\cite{paterek2007improving} and \texttt{SVD++}\cite{Koren:2008}; Cornerstone ranking oriented algorithms under this paradigm are \texttt{BPR} \cite{BPR}, \texttt{WRMF}\cite{hu2008collaborative}, \texttt{CDAE}\cite{Wu2016CollaborativeDA}.

\emph{Multi-task learning} (\texttt{MTL}) ~\cite{caruana1998multitask} deals with problems where multiple tasks with some commonalities are given. The learning procedure optimizes these tasks simultaneously, using representation sharing. Due to the common traits of the tasks, such an optimization method can improve generalization and accuracy for the task-specific models, when compared to training the models separately.

\section{Related Work} \label{sec:related}
\texttt{MTL} has been shown to be beneficial in variety of tasks, including the problem of generating personalized recommendations, that we focus here. However, the research on this problem is still sparse. The first work in this area (\cite{shi2013unifying}) suggested training simultaneously a ranking and a rating prediction algorithms with shared embedding matrices. The ranking algorithm was chosen to be \texttt{ListRank} \cite{shi2010list} and the rating algorithm was Probabilistic Matrix Factorization \cite{mnih2008probabilistic}. Later, the authors of \cite{li2017one} proposed to use \texttt{CLiMF} \cite{shi2013climf} as the ranking algorithm and reported improved results.
These works are conceptually similar, as they differ only in the choice of the concrete selection of the rating prediction algorithm. One can unify these works in a framework that uses the \emph{same} representation for both learning tasks, i.e., the same user and item vectors are used for ranking the items and for predicting the actual ratings.

Another line of work incorporates both implicit feedback and explicit ratings for either ranking or rating prediction tasks. A landmark algorithm that falls under this umbrella is \texttt{SVD++}\cite{Koren:2008}, which solves the rating prediction task. The user representation generated by this algorithm considers all items for which the user made an implicit feedback. It further considers all explicit ratings also as implicit feedback interactions. 

On the other side, several ranking-oriented algorithms were proposed, that integrate explicit ratings. For instance, \texttt{BPR} \cite{BPR} is a seminal ranking algorithm. The algorithms presented in \cite{pan2015adaptive,wang2012please} enhance it by allowing it to support explicit ratings. 
For last, the work presented in \cite{li2016exploiting} integrates the rating prediction algorithm \texttt{SVD++} with the ranking algorithm \texttt{xCLiMF}\cite{shi2013xclimf}.
These types of work, however, do not serve as a general framework that unifies ranking and rating prediction tasks. Instead they are tailored to specific algorithms and generalizing them to any other algorithm may not be trivial.

\section{Our Approach} \label{sec:ours}
In this work, we argue that the ranking and rating prediction tasks are not completely separate problems, but are part of a single process. We accompany the description of our intuition with examples from the movies domain, although the presented concepts are general and can be applied to other domains as well. First, the user decides to interact with a specific item (watch a movie). The ranking task, which is also known as the selection task, aims at forecasting this choice of the user.
The decision made by the user is based on several weighted criteria. For example, the user may decide to watch only Hollywoodian movies and highly prefers comedies.
After interacting with the item, in post-consumption perspective, the user becomes more familiar with the item's traits, may notice nuances and her perception of the item is modified. In our running example, the user may understand that the movie is funnier than expected.
Then the user may leave an explicit rating that summarizes the assessment of the user toward the item. At this point, the user's system of considerations is different comparing to the previous choice. For instance, given that the user has watched the movie, the assigned rating may be heavily influenced by the plot, and less affected by whether the movie was shot in Hollywood. The rating prediction task is aligned with this decision.

We therefore design a general framework that unifies the ranking and rating prediction tasks while supporting the described two-phase decision process of ranking and rating. We hereinafter dub the proposed algorithm as ``Rank and Rate'', or \texttt{RnR} for short.
Let $(\pazocal{R},\pazocal{P})$ be a pair of any latent factor models that solve the ranking and the rating prediction tasks correspondingly. These algorithms will serve as the underlaying algorithms for \texttt{RnR}. Our proposed approach is a general framework, and is agnostic to the concrete choice of the underlaying algorithms, as long as they can be optimized by Stochastic Gradient Descent based optimization methods. They can even vary in the input, required at training time, or to follow different optimization approaches (i.e., point-wise, pair-wise or list-wise). For instance, a possible choice of ranking algorithm can be \texttt{WRMF}\cite{hu2008collaborative}, which requires only the identities $u,i$ of the target user and item. Another natural candidate for the ranking algorithm is \texttt{BPR}\cite{BPR}, which in a addition to the user and item identities, requires a negative sampled item. For the rating prediction algorithm one can consider \texttt{SVD}\cite{koren2009matrix} or \texttt{SVD++}\cite{Koren:2008}. While the former requires only the identities $u,i$ and the explicit rating, the latter also requires the set of all implicit feedbacks given by the user $u$.

\begin{figure}[t]
\begin{center}
\includegraphics[width=0.6\linewidth]{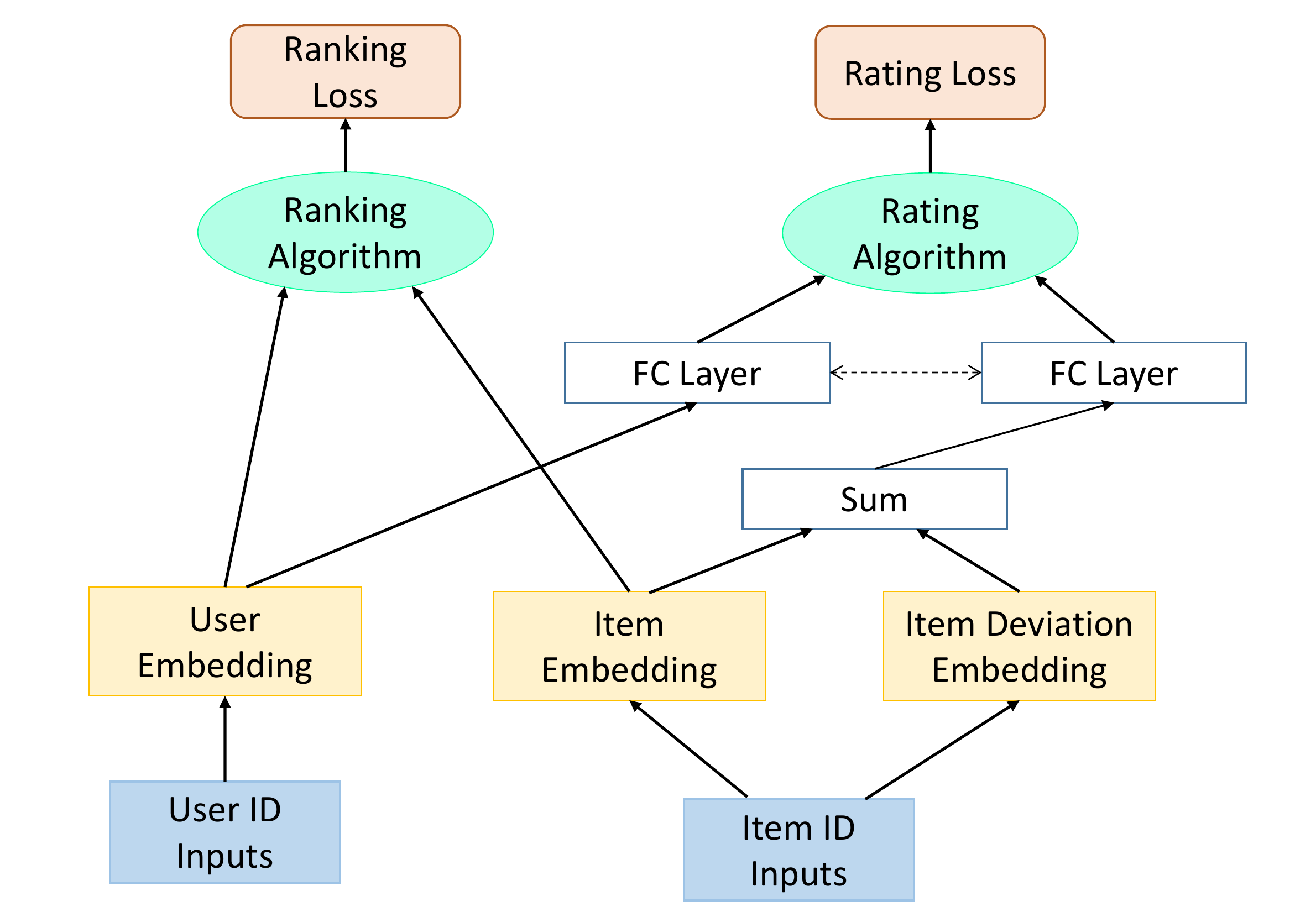} 
\caption{Model Architecture - illustrates the two-phase process, where we first rank using the raw embeddings, and then rate using the transformed embeddings.}
\label{fig:Architecture}
\end{center}
\end{figure}

Let $L_P(\pazocal{P},D;U,I)$ be the loss achieved by algorithm $\pazocal{P}$ on dataset $D$ with user and item embedding matrices $U$ and $I$\footnote{We assume a single item embedding matrix, but it is straightforward to support multiple matrices}. Similarly, we define the loss of the ranking algorithm $\pazocal{R}$ as $L_R(\pazocal{R},D;U,I)$.
\texttt{RnR} performs multi-task learning, and optimizes the parameter sets of $\pazocal{R}$ and $\pazocal{P}$ simultaneously, while sharing the embedding matrices $U$ and $I$. Unlike previous methods, our approach applies the two-phase decision process by modifying the user and item representations for the \emph{rating prediction task}. Namely, we iterate over the dataset $D$ and simultaneously optimize both $\pazocal{R}$ and $\pazocal{P}$. While $\pazocal{R}$ uses the raw embedding matrices $U$ and $I$, $\pazocal{P}$ uses parameters that support the second decision made by the user. Specifically, we learn an item deviation matrix $I^d$  which models a post-consumption perspective of the users toward items. For each item $i$, it gives us a deviation  item vector $q_i^d$, which we add to the original  item vector $q_i$ to obtain the modified item vector $q^{post}_i$. Therefore, $q^{post}_i$ reflects the post-consumption perception of the item. We perform an unweighted sum  $q^{post}_i=q_i+q_i^d$, since the parameters in $I^d$ can learn the relative importance of each component in the sum.  Next, we map the user and the updated item vectors to a new feature space which reflects the post-consumption decision process of the user. We implement the mapping by a fully connected layer (FC)  with a non-linear activation function.  We tie the parameters of the user and item FC layers, hence the new representations $p_u^P$ and $q_i^P$ of users and items for the rating prediction task are projected to the same latent space. Formally, $p_u^P=FC_\theta(p_u), q_i^P=FC_\theta(q_i^{post})$, where $FC_\theta(\cdot)$ denotes the output of a fully connected layer parameterized by $\theta$. The architecture of \texttt{RnR} is shown in Figure~\ref{fig:Architecture}.

Our proposed framework mimics the decisions making process as follows. We go over all tuples $(u,i,r)\in D$. For each, we first invoke $\pazocal{R}$ to predict the ranking choice. We then invoke $\pazocal{P}$ with the modified parameters as explained above, to predict the rating choice.  We define the objective function as follows,
\begin{equation}
\begin{split}
O = \min_{U,I,I^d,\theta}  {} &\alpha\cdot L_R(\pazocal{R},D;U,I) \\ 
& + (1-\alpha)\cdot L_P(\pazocal{P},D;FC_\theta(U),FC_\theta(I+I^d)) \\
& + \lambda(\|U\|^2+\|I\|^2+\|I^d\|^2+\|\theta\|^2)
\end{split}
\end{equation}
where $\alpha$ and $\lambda$ are hyper-parameters that control the relative contribution of each individual task and the regularization factor, respectively.	

\section{Results} \label{sec:results}
In our experiments, we test the performance of proposed \texttt{RnR} method, which combines the multi-task learning of ranking and rating prediction tasks with the two-phase decision process. We compare it to two types of baselines: 1) a vanilla multi-task framework that learns these two tasks simultaneously, but without the two-phase decision process\cite{shi2013unifying,li2017one} and 2) single-task algorithms:  Collaborative Denoising Auto-Encoders\cite{Wu2016CollaborativeDA}, Bayesian Personalized Ranking\cite{BPR}, and \texttt{SVD}\cite{paterek2007improving}. We also report the results of Popularity, which is always predicting the top k most frequent items in train, for completeness.

For the vanilla multi-task learning, given the underlaying algorithms for ranking and rating prediction, $\pazocal{R}$ and $\pazocal{P}$, we optimizes the parameters using the combined objective:
\begin{equation}
\begin{split}
O = \min_{U,I}  {} &\alpha\cdot L_R(\pazocal{R},D;U,I) + (1-\alpha)\cdot L_P(\pazocal{P},D;U,I) \\ 
& + \lambda(\|U\|^2+\|I\|^2).
\end{split}
\end{equation}
As in \texttt{RnR}, the role of $\alpha$ is to control the relative contribution of each individual task.

For ranking, we choose two different algorithms: Collaborative Denoising Auto-Encoders\cite{Wu2016CollaborativeDA} and Bayesian Personalized Ranking\cite{BPR}. For rating prediction we use \texttt{SVD}\cite{paterek2007improving} (the algorithms are detailed below).

\subsection{Underlaying Algorithms}
\emph{Collaborative Denoising Auto-Encoders} (\texttt{CDAE})~\cite{Wu2016CollaborativeDA} is an encoder-decoder architecture, that achieves state-of-the-art results on the personal recommendations task. It is consists of three main concepts. First, is the \emph{Auto-Encoder}, which is given as input a set of items a user has interacted with. Then it generates an internal representation of the user (the so-called user vector), which in turn allows to reconstruct the original input. The second concept, \emph{Collaborative}, means that the identity of the user is also given as input, which allows to refine the internal representation of the user. The last concept \emph{Denoising} improves the generalization of the model by asking to reconstruct the complete historical usage from a partially observed list of items. 

\emph{Bayesian Personalized Ranking} (\texttt{BPR}) method~\cite{BPR} is a simple, yet effective and widely adopted implicit feedback MF method. \texttt{BPR} is optimized for correctly ranking observed user-item interactions over non-observed ones.

\texttt{SVD}\cite{paterek2007improving} is a seminal rating prediction algorithm. It factorizes a usage matrix (with explicit feedback) into two low rank matrices, for users and items. This distributed representation, together with learned user and item biases, is learned to minimize the squared error of the reconstructed ratings.

\subsection{Datasets}
Our evaluation is done on two real-world datasets, from different domains.

\textbf{MovieLens}\footnote{\small \url{http://grouplens.org/datasets/movielens/1m/}}: This dataset contains about 1 million ratings that were applied to more than 3,700 movies by more than 6,000 users.

\textbf{Yelp}:\footnote{\small \url{http://www.yelp.com/dataset/challenge}} This is the most recent dump of the Yelp challenge. It contains about 3.5M reviews on about 170,000 businesses made by about 295,000 users. 

Each user interaction in these datasets is associated with an explicit rating on a 1-5 Likert. Users with less than 4 interactions were omitted.

\subsection{Evaluation Protocol and Experiments}
Using each dataset, we evaluated our approach and the baselines using an ``All But (Last) One'' protocol (i.e., the hidden item belongs to the chronologically last interaction), namely, we remove the last rated item for $n$ users from the training set, and put half of the removed items in the validation set and the other half in the test set. We used  $n=5,000$ in the  MovieLens experiments and  $n=50,000$ in Yelp experiments. The results are reported on Recall@k (denoted by Accuracy@k) and MRR@k \cite{shani2011evaluating}, for k=10.

We implemented our and baseline methods using TensorFlow \footnote{\small \url{https://www.tensorflow.org/}} employed with the AdaGrad optimizer \cite{duchi2011adaptive}, with learning rate of 0.001. User and item embedding size was fixed to be 50.
We did modest grid search over  $\alpha \in [0.9, 0.95, 1.0]$ and regularization $\lambda \in [0.01, 0.001]$.

We initialized the deviation matrix $I^d$ and biases with zeros, since its used in summation with other embeddings or scalars. All other parameters were initialized using Xavier initialization. 

\subsection{Results}
The complete list of results is reported in Table \ref{tab:results}. We  first observe that CDAE outperforms BPR, and in general BPR achieves better results than SVD. Second, combining ranking and rating algorithms using the vanilla approach always improves over single-task algorithms. We can further note that our method significantly improves the vanilla approach. This is attributed to the support in the two-phase decision process made by users.

\begin{table}
\begin{center}
% \resizebox{.6\columnwidth}{!}{%
\begin{tabular} { |c||c|c||c|c| } 
 \hline
  & \multicolumn{2}{c||}{MovieLens 1M} & \multicolumn{2}{c|}{Yelp 2018} \\ 
  Algorithm &   Accuracy  & MRR & Accuracy  & MRR \\ 
 \hline
 Popularity & 0.0278 & 0.0035 & 0.0166 & 0.0053 \\ 
 \hline
 \texttt{SVD} & 0.0673 & 0.0120 &  0.0043 & 0.0010 \\ 
 \hline
 \texttt{CDAE} & 0.0926 & 0.0146 & 0.0305 & 0.0107 \\ 
 Vanilla(\texttt{CDAE}, \texttt{SVD}) & 0.1040 & 0.0189 & 0.0318 & 0.0115 \\ 
 \texttt{RnR}(\texttt{CDAE}, \texttt{SVD}) & \textbf{0.1236} & \textbf{0.0497} & \textbf{0.0400} & \textbf{0.0137} \\ 
 \hline
 \texttt{BPR} & 0.0659 & 0.0222 & 0.0196 & 0.0071 \\ 
 Vanilla(\texttt{BPR}, \texttt{SVD}) & 0.0732 & 0.0244 & 0.0207 & 0.0054 \\ 
 \texttt{RnR}(\texttt{BPR}, \texttt{SVD}) & \textbf{0.0752} & \textbf{0.0263} &  \textbf{0.0238} & \textbf{0.0077} \\ 
 \hline
\end{tabular} %
% }
\end{center}
\caption{Experiments results, higher is better.}
\label{tab:results}
\end{table}

\section{Conclusions} \label{sec:conclusions}

In this work, we showed a generic, yet efficient way of improving a recommendation system which combines multi-task learning of ranking and rating prediction with two-phase decision process made by users. For the latter, we used an explicit rating information, which is available in many cases.  We showed that the proposed framework can combine different underlying algorithms, improving them significantly in all cases. Hence, a practitioner can use the proposed framework with any other ranking  and raring  predictions methods.

We expect that the results of this work will inspire further research on using the explicit signal in efficient ways.
\clearpage
\bibliographystyle{alpha}
\bibliography{sample}

\end{document}